\documentclass[aps,prd,nofootinbib,amsmath,amssymb,twocolumn]{revtex4-2}
\usepackage[utf8]{inputenc}
\usepackage{xcolor}
\usepackage{graphicx}
\usepackage{verbatim}
\usepackage{mathrsfs}
\usepackage{orcidlink}
\usepackage{hyperref}

\newcommand{\rd}{\textrm{d}}

\begin{document}

\title{The G\"odel Universe as a Superconductor}

\author{Daniel~\surname{Flores-Alfonso}~\orcidlink{0000-0001-7866-3531}}
\email[]{dafa@azc.uam.mx}
\affiliation{Departamento de Ciencias B\'asicas, Universidad Aut\'onoma Metropolitana -- Azcapotzalco, Avenida
San Pablo 420, Colonia Nueva El Rosario, Azcapotzalco 02128, Ciudad de M\'exico, Mexico}

\author{Cesar S.~\surname{Lopez-Monsalvo}~\orcidlink{0000-0002-0378-0415}}
\email[]{cslm@azc.uam.mx}
\affiliation{Departamento de Ciencias B\'asicas, Universidad Aut\'onoma Metropolitana -- Azcapotzalco, Avenida
San Pablo 420, Colonia Nueva El Rosario, Azcapotzalco 02128, Ciudad de M\'exico, Mexico}

\author{Alberto~\surname{Rubio-Ponce}~\orcidlink{0000-0002-7562-2918}}
\email[]{arp@azc.uam.mx}.             
\affiliation{Departamento de Ciencias B\'asicas, Universidad Aut\'onoma Metropolitana -- Azcapotzalco, Avenida
San Pablo 420, Colonia Nueva El Rosario, Azcapotzalco 02128, Ciudad de M\'exico, Mexico}

\begin{abstract}
Material science and engineering have benefited from the use of geometric and topological tools. A material medium  can mimic effective gravitational fields while spacetime metrics serve as geometric models of physical media. Although analog models of optical, acoustic, and viscous media in curved spacetimes are well established, none have yet captured the hallmark constitutive properties of superconductors. In this work we show that the G\"odel universe -- an exact solution to Einstein's field equations -- serves  as the gravitational analog of a superconducting medium in its Meissner state. 
\end{abstract}

\maketitle

Research in electromagnetic materials spans solid state physics, optics and  engineering, increasingly incorporating  tools stemming from general relativity to design  metameterials~\cite{Leonhardt:2000fd,Leonhardt:2006ai,Rahm:2008,Li:2008}. This builds on the century-old proposal of  spacetime as an effective medium with position dependent refractive index ~\cite{Gordon:1923qva}. While much attention has focussed  on nonconductive materials, recent work has shown that certain  (2+1)-dimensional spacetimes provide us with an equivalent geometric framework to that of London equations governing superconducting currents ~\cite{Flores-Alfonso:2020zif,Lopez-Monsalvo:2023egs}.  These equations, alongside Maxwell's, describe the macroscopic response of superconductors to external electromagnetic fields~\cite{London:1935} and can be formulated in the language of differential forms~\cite{Sternberg:2012}, allowing us to extend the analysis to higher dimensions and richer geometric configurations. 

In this paper, we reveal that  G\"odel's 1949 solution to Einstein's equations~\cite{Godel:1949ga} -- a homogeneous, rotating universe filled with a perfect fluid --  models a superconducting medium under an external electromagnetic field. Here,  the superconducting current aligns with the free falling observers' flow, while the constitutive relation establishes its dual nature with the electromagnetic potential, satisfying the Klein-Gordon equation with an effective mass term given by the underlying spacetime metric. This establishes the G\"odel universe as a unique four-dimensional superconductor analog, bridging gravitational physics with condensed matter and opening new paths for material science exploration via general relativity.

Let us start by writing the G\"odel metric as~\cite{Godel:1949ga,Hawking:1973uf}
\begin{align} \label{metric}
    \rd s^2 = &-\left[\rd t +\frac{2}{\Omega} \sinh^2{\left(\frac{\sqrt{2}\Omega r}{2}\right)} \ \rd \phi\right]^2 + \rd r^2 \notag \\
    &+\frac{1}{2\Omega^2} \sinh^2{\left(\sqrt{2}\Omega r\right)}\rd \phi^2 + \rd z^2,
\end{align}
with $\Omega$ a constant. A simple computation shows that this metric solves the Einstein field equations in the presence of a perfect fluid,
\begin{equation}
\label{EFE}
    G = (\rho+p)\ u \otimes u+p\ g,
\end{equation}
whenever $\rho=p=\Omega^2$ and 
\begin{equation} \label{u}
    u = \rd t +\frac{2}{\Omega} \sinh^2{\left(\frac{\sqrt{2}\Omega r}{2}\right)}\rd \phi.
\end{equation}
Thus, the matter filling the universe is a stiff fluid satisfying the equation of state $\rho=p$, and flows along the direction $u$. Notice that observers comoving with the fluid experience rigid rotation as their corresponding worldlines have vanishing expansion and shear, indicated by
\begin{equation}
    \nabla u = -\frac{1}{2}\rd u.
\end{equation}
What is more, their vorticity is constant,
\begin{equation}
    (\nabla u)^2=2\Omega^2,
\end{equation}
meaning the fluid swirls around the universe at a constant rate.

The fluid's constant energy density is a reflection of the high symmetry present in the G\"odel universe. In fact, the spacetime is a homogeneous space with a five-dimensional transitive isometry group. The metric has three evident symmetries which are characterized by the Killing vector fields $\xi_1 = \frac{\partial}{\partial t}$, $\xi_2 = \frac{\partial}{\partial \phi}$, and $\xi_3 = \frac{\partial}{\partial z}$ and two additional symmetries described by
\begin{subequations}
\begin{align}
\xi_4 &= \frac{1}{\Omega}\tanh{\left(\frac{\sqrt{2}\Omega r}{2}\right)}\sin\phi\frac{\partial}{\partial t}-\frac{\sqrt{2}\cos\phi}{2\Omega}\frac{\partial}{\partial r}+\notag \\
&\phantom{=}~\coth{\left(\sqrt{2}\Omega r\right)}\sin\phi\frac{\partial}{\partial \phi}, \\
\xi_5 &= \frac{1}{\Omega}\tanh{\left(\frac{\sqrt{2}\Omega r}{2}\right)}\cos\phi\frac{\partial}{\partial t}+\frac{\sqrt{2}\sin\phi}{2\Omega}\frac{\partial}{\partial r}+\notag \\
&\phantom{=}~\coth{\left(\sqrt{2}\Omega r\right)}\cos\phi\frac{\partial}{\partial \phi}.
\end{align}
\end{subequations}
It can be shown that the G\"odel spacetime is the only perfect fluid solution with such a high degree of symmetry. Consequently, it is a unique system which exhibits this combination of homogeneity, rigidity, and equation of state.

Notice that the G\"odel spacetime has a direct product structure, and the metric \eqref{metric} can be written as $\rd s^2 = \rd s^2_{(3)}+\rd z^2$. Hypersurfaces for constant values of $z$, say $z_0$, are three-dimensional homogeneous spaces with warped anti-de Sitter (WAdS) geometry~\cite{Rooman:1998xf}. A remarkable property of WAdS$_3$ is that it is a contact manifold. A convenient parametrization of its contact structure is provided by
\begin{equation}
    \eta = u|_{z=z_0}.
\end{equation}
Contact forms such as $\eta$ satisfy the nonintegrability condition $\eta\wedge\rd\eta\neq0$. They also determine a distinguished direction given by the Reeb vector field $\xi$ which is defined by the conditions $\rd\eta(\xi)=0$ and $\eta(\xi)=1$. In the present case, the Reeb vector is timelike and is aligned with the direction of fluid flow in the G\"odel spacetime. In other words, the vorticity of the fluid velocity is, to some extent, a realization of the underlying contact structure.

Let us remark that geometric conditions exist in order to establish if a given three-dimensional metric is a solution to Einstein equations together with a perfect fluid or an electromagnetic field~\cite{Krongos:2015fta,Krongos:2016bqp}. In order to include fields with a Chern-Simons term we refer to Appendix B of \cite{Flores-Alfonso:2023fkd}. For instance, one could apply these purely geometric procedures to WAdS$_3$ metrics and derive the solutions of~\cite{Banados:2005da}, constructing the matter content from the metric as determined by the algorithm. For the G\"odel metric, the four-dimensional perfect fluid conditions~\cite{Krongos:2015fta} are obviously satisfied. Nonetheless, it fails to satisfy the Rainich conditions~\cite{Rainich:1925,Torre:2013nia} which would establish it as an electrovacuum solution. It is also not a charged dust solution, despite the fact that when $z=z_0$ it does admit that interpretation. Having made these last points, we move on to our main result: portraying the G\"odel universe as a superconductor, for which we must apply to the geometric model an external electromagnetic field.

Even though the G\"odel spacetime is not sourced by any electromagnetic field we are still able to ask ourselves what is the most general field compatible with its symmetries. From the outset, notice that any field compatible with the symmetries must necessarily take the form
\begin{equation} \label{field}
    F = \frac{B}{\sqrt{2}\Omega}\sinh{\left(\sqrt{2}\Omega r\right)}\rd r \wedge \rd \phi,
\end{equation}
where $B$ is an integration constant with dimensions of magnetic flux density. Notice that a comoving observer determines the magnetic flux density in terms of the $B$ parameter, that is,
\begin{equation}
\dot\iota_{u^\sharp} \star F = B\ \rd z, \quad {\rm where } \quad u^\sharp = g^{ab} u_b \frac{\partial}{\partial x^a} = \frac{\partial}{\partial t}.
\end{equation}
It is perhaps unsurprising that the only electromagnetic field invariant under the G\"odel symmetries is proportional to the curl of the fluid velocity
\begin{equation}
    F=\frac{B}{2\Omega}\rd u.
\end{equation}

Our hypothesis is that the G\"odel geometry completely characterizes how the medium it models responds to external fields. Notice that when the field in Eq. \eqref{field} is applied a current is induced on the material, thus establishing it as a conductor. This follows from the fact that
\begin{equation}
   \star \rd \star F = j,
\end{equation}
for an induced current 
\begin{equation}
    j = -2\Omega B\,u.
\end{equation}
Observe that the charge density $\sigma=2\Omega B$ is constant and that charge circulates along fluid flow in the G\"odel solution, cf. Eq. \eqref{u}. Since the induced current is aligned with the fluid velocity and Eq. \eqref{EFE} has vanishing heat flux~\cite{Eckart:1940te} this suggests negligible electric dissipation, akin to that of a perfect conductor. However, the existence of true superconductivity would require further evidence of flux quantization or a phase transition. Before considering such questions we pursue a more basic inquiry and immediately check if the induced current and applied field satisfy the London constitutive equations \cite{London:1935,Sternberg:2012}
\begin{equation}
    j= -\frac{1}{\lambda^2}A,
\end{equation}
finding that they are fulfilled whenever $\lambda=1/2\Omega$. These equations describe the Meissner effect that takes place in superconductors for which $\lambda$ is the London penetration depth. In other words, when a uniform magnetic field is applied the G\"odel universe responds as a superconducting medium. Notice that conservation of the electric current $\rd\star j=0$ together with the London equation above imply that $\rd\star A=0$. Moreover, when the Maxwell and London equations are combined they yield $\star \rd \star F=-\lambda^{-2}A$. This is to say, after the superconducting electric current is induced the gauge field acquires an effective mass given by the metric's vorticity parameter
\begin{equation}
    \Delta A+(2\Omega)^2A =0,
\end{equation}
where $\Delta=\delta\rd+\rd\delta$ is the Laplace-de Rham operator. Alternatively, by applying the Weitzenb{\"o}ck identity~\cite{Choquet-Bruhat:1982,Barrientos:2019msu} we are able to rewrite this expression as the wave equation 
\begin{equation}
    \Box A-R_{ab}A^{a}\rd x^{b}-(2\Omega)^2A =0,
\end{equation}
where $\Box=\nabla^{\mu}\nabla_{\mu}$ is the Laplace-Beltrami operator. Furthermore, on the G\"odel background the curvature-coupling terms simplify and yield
\begin{equation}
    \Box A-2\Omega^2A =0.
\end{equation}

Moreover, as observed in lower dimensions~\cite{Lopez-Monsalvo:2023egs}, the superconducting current flows along geodesics of the model metric since it is transverse to the field $\dot\iota_{u^\sharp} F =0$. This is to say, since the field is force free then the superconducting current has a Beltrami flow. Consequently, the current does not flow along any of the iconic closed timelike curves of the G\"odel universe, as such curves are never geodesic~\cite{Kundt:1956,Chandrasekhar:1956}.

A complementary point here is that spacetime geometries are renowned for acting as effective dielectric media~\cite{Plebanski:1959ff}, in particular as impedance-matched anisotropic media~\cite{Leonhardt:2006ai}. For the Hodge constitutive relation $H=\star F$, the components of the excitation field $H$ relate to those of the field strength $F$ by
\begin{equation}
    H_{ab} = \frac{1}{2}\sqrt{-g}\epsilon_{abef}\chi^{efcd}F_{cd},
\end{equation}
where
\begin{equation}
    \chi^{abcd}=\frac{1}{2}\left(g^{ac}g^{bd}-g^{ad}g^{bc}\right).
\end{equation}
These expressions are convenient for calculating the properties of a material that is analog to a curved spacetime. In particular, the medium's permittivity and permeability are obtained from~\cite{Hehl:2003,Obukhov:2004zz,Schuster:2018cwt}
\begin{align}
    \varepsilon^{ab} &= -2\chi^{acbd}u_{c}u_{d},\\
    [\mu^{-1}]_{ab} &= \frac{1}{2}|\det{g}|\epsilon_{acde}\epsilon_{bfgh}\chi^{degh}u^{c}u^{f},
\end{align}
respectively. For the G\"odel universe, written as (0,2) tensors, these become
\begin{equation}
    \varepsilon=[\mu^{-1}] = \rd r^2+\frac{1}{2\Omega^2} \sinh^2{\left(\sqrt{2}\Omega r\right)}\rd \phi^2 + \rd z^2.
\end{equation}
Notice that the medium has some degree of homogeneity and isotropy inherited from the spacetime symmetry, whereas it is fully isotropic optically. As such, the analog material is compatible with a structure that has axial symmetry. The physical boundary of the material can be made to match the optical horizon (see Ref.~\cite{Kundt:1956}) of the spacetime origin.

Let us conclude by mentioning that this work was motivated by 3D spacetimes on which electric currents along the Reeb vector and electromagnetic fields aligned with the contact structure satisfied the (2+1)-dimensional version of the London equations~\cite{Flores-Alfonso:2020zif}. Even for product spaces like the G\"odel universe it is not obvious \emph{a priori} that the London or Maxwell equations should hold under similar conditions. To address this issue, we have first examined how many electromagnetic fields may exist on the G\"odel background compatible with its symmetries and found that there is only one such field, and it is given by Eq. \eqref{field}. Our results show that a material whose response functions are modeled by the G\"odel metric exhibit superconducting behavior under a ``uniform'' electromagnetic stimulus. In particular, the induced current on the material experiences no electrical resistance and a vanishing Lorentz force. At the same time, the applied field and induced current obey the London constitutive relations. Our approach demonstrates that the toolbox of general relativity is also useful for modeling conducting materials, complementing the practical value it is renowned for in metamaterials, especially in transformation optics~\cite{Leonhardt:2006,Pendry:2006}.

Evidently, since the G\"odel universe is a homogeneous space and the applied field is invariant under all its isometries then it is not surprising that the charge density $\sigma$ of the superconducting current is constant in this case. As such, our example represents a four-dimensional analog of the geodesic induced currents analyzed in~\cite{Lopez-Monsalvo:2023egs}. It would be interesting to fully extend the approach taken there to four dimensions as well as examine geodesic currents with more general charge densities and their associated geometry, especially in light of the solutions found in~\cite{Corral:2024xfv} where rich and diverse gauge systems were found beyond the Abelian theory. This suggests that it will prove valuable to further understand nonintegrable structures in gauge theory, in particular, if this line of inquiry ultimately leads to novel developments in color superconductivity. On this basis, the results of this work serve as the foundation for future findings.

Lastly, we point out that in the weak field limit where $B/\Omega\ll1$ the fluid in the G\"odel universe is only slightly charged, meaning the magnitude of its charge density is much less than its energy density $|\sigma|\ll\rho$. Then the circulating electric current would sustain the magnetic field in Eq. \eqref{field}, and the matter equations are solved exactly. Moreover, the Einstein equations are solved approximately in this case, since the Maxwell energy-momentum tensor is negligible when compared to that of the stiff fluid. Thus, as a side product our analysis has led to a new approximate charged perfect fluid solution~\cite{Synge:1960ueh} that uses the G\"odel metric as background. Furthermore, since the electromagnetic field is compatible with all of the background isometries the charged system retains its original homogeneity. Remarkably, because the magnetic field is transverse to the electric current the Lorentz force is vanishing, and so the system remains dynamically stable. It is worth noticing that this type of electromagnetic behavior is not unique of this particular solution. Indeed, our conclusion can be directly verified in the Som-Raychaudhuri spacetime~\cite{Som:1968}, which exactly solves the charged dust Einstein-Maxwell equations.

Succinctly, our work presents an explicit example of the nonintegrable geodesic flow characteristic of superconducting currents where the medium is an exact solution to Einstein field equations.

D.F.A. would like to acknowledge financial support from SECIHTI through a postdoctoral research grant.

\bibliographystyle{apsrev4-2}

\end{document}